\documentclass[a4paper,man,natbib]{apa6}
\usepackage{setspace} 

\usepackage{subcaption}

\usepackage[english]{babel}
\usepackage[utf8x]{inputenc}
\usepackage{amsfonts,amsmath,amssymb}
\usepackage{mathtools}
\usepackage{graphicx}
\usepackage[colorinlistoftodos]{todonotes}
\usepackage{comment}
\usepackage{booktabs}
\usepackage{csquotes}
\usepackage{lipsum}
\usepackage{tikz-cd}

\newtheorem{theorem}{Theorem}

\newtheorem{definition}[theorem]{Definition}

\usepackage{comment}
\usepackage{soul}

\title{Theoretical Note: On the Practical Uses of Mathematical Theory for Attitude Research}
\shorttitle{Attitudes and Graph Dynamical Systems}

\author{Mark G. Orr\textsuperscript{1,}\textsuperscript{4}, Emily S. Teti\textsuperscript{2}, Andrei Bura\textsuperscript{3} and Henning Mortveit\textsuperscript{3}}

\affiliation{
\textsuperscript{1}Florida Institute for Human and Machine Cognition\\ 
\textsuperscript{2}Los Alamos National Laboratory, Nuclear Engineering and Nonproliferation Division\\
\textsuperscript{3}University of Virginia, Biocomplexity Institute\\
\textsuperscript{4}Corresponding Author; morr@ihmc.org}

\abstract{In attitude theory, formal theoretical predictions come largely from the simulation of computational models.  We argue that to push attitude theory further, we should employ mathematical analysis/analytic methods alongside of computational simulation, something that other sciences and engineering consider standard practice.  Our work first attempts to portray the complementary nature of mathematical analysis along side of computational simulation using as an example the Causal Attitude Network model of attitudes \citep{dalege2016}.  We then introduce a mathematical theory, Graph Dynamical Systems (GDS), as a broad theoretical framework for network models of attitudes.   We illustrate the use of GDS, in the context of the Attitudes as Constraint Satistfaction (ACS) theory of attitude dynamics \citep{MonroeRead2008}, as a generator of precise, quantitative theoretical predictions.  We conclude by pointing out the value of improved attitude theory for the so-called replication crisis in psychology. 
}

\AtBeginDocument{\singlespacing}

\begin{document}
\maketitle


\section{Introduction}
In social psychology in general and in attitude theory in particular, computational modeling carries the burden of formal theory development. This stands to reason.  Computational modeling provides: (i) clear definition of theoretical components and their relations, (ii) by (i) clear explanation of psychological phenomena, and, (iii) quantitative theoretical predictions via simulation of mental and social process for purposes of empirical validation. But, the reliance on computational modeling for theory development leaves mathematical theory underdeveloped. As theories of attitudes are accumulating sophisticated theoretical concepts and predictions (we provide examples shortly), we think it high time to incorporate mathematical theory that can clearly articulate sophisticated theoretical predictions and complement the predictions that are derived from existing computational simulation approaches.  

In this article, we offer a mathematical theory for the study of attitude dynamics.\footnote{Early notable efforts are \citep{cartwright1956generalization,AjzenFishbein1975,hunter2014mathematical}.}  Our efforts leverage prior work on computational approaches to attitudes, efforts that put some useful mathematical machinery in place \citep[e.g.][]{MonroeRead2008,OrrThrushPlaut2013,dalege2016}.  For the study of attitudes, the primary purpose of mathematical theory is the same as for computational formalisms: to provide (i) explanation and (ii) novel theoretical predictions for empirical validation in the service of theory development. 

The course of our article is as follows.   First, we illustrate the value of mathematical theory in proffering complement to simulation approaches using a simple prototypical example from the attitude literature (the Causal Attitude Network and Attitudinal Entropy framework \citep{dalege2016,DalegeBorsboom_2018}).  Subsequently, we introduce our mathematical theory, Graph Dynamical Systems, and furnish an example of its application to the Attitudes as Constraint Satisfaction Theory \citep{MonroeRead2008}.  Computational modeling and simulation approaches are not in competition with mathematical theory; they are complementary.  Our primary thesis, thus, is that co-opting mathematical theory may benefit theoretical development in social psychology in general and attitude research in particular. 
\subsection{Computational Theories of Attitudes}
Over the past 25 years, the social psychological literature has successfully integrated some of the computational modeling approaches from cognitive science to address a range of phenomena:  causal attribution, stereotypes, attitude formation, impression formation, and personality \citep[e.g.][]{ReadMiller1998,Vallacher2017,Overwalle2007,Smith1996,OrrThrushPlaut2013,MonroeRead2008,Conrey2007attitude}.   Constraint satisfaction, typically formalized and implemented as a kind of recurrent artificial neural network, holds a strong position in social psychology.  Not only does it fit past and current understandings of a range of phenomena, it offers a formalization of mechanism \citep[see][]{ReadVanmanMiller1997, Simon2002structural}. Computational models of attitude formation and change are dominated by the constraint satisfaction formalism \citep[e.g.][]{OrrThrushPlaut2013,MonroeRead2008,orr2014complex,van2005connectionist,Conrey2007attitude,ehret2015modeling,overwalle2005a}. 

Over the past decade, constraint satisfaction models and associated attitudinal theory have been re-invigorated in social psychology with the introduction of a new class of attitude model.  The Causal Attitude Network (\textbf{CAN}) model and close-cousin Attitudinal Entropy (\textbf{AE}) framework, have been cast, in their union, as a novel theoretical approach for understanding attitude formation and change \citep{dalege2016, DalegeBorsboom_2018, DalegevanderMaas2020}\footnote{It is referred to as "new."\citep[see p. 3, col 1, last paragraph]{dalege2016}}.  This approach stems from the psychological networks approach that rose to prominence in the clinical psychology literature over the past decade or so \citep{Bringmann2018, Borsboom2017, BorsboomCramer2013, Bringmann2021, Borsboom2008, CramerWaldorpMaasBorsboom2010}.  For purposes of this article, we will dub this class of attitude model as the CANAE approach or framework or, for simplicity, just CANAE.

The novelty of CANAE, on the surface, stems from: 
(i) its use of constructs from statistical physics (e.g., Gibbs and Boltzmann distributions of the configuration space, pseudo-thermodynamic temperature effects), 
(ii) its use of constructs from network science (e.g., global and local topological properties of the networks)\footnote{Under the umbrella of \textit{psychological networks}}, and 
(iii) its learning mechanism. 

CANAE is an Ising-like model \citep{dalege2016} and, thus, falls within the larger class of constraint satisfaction models of attitudes. Others have used statistical physics counter-parts in constraint-satisfaction models of attitude, e.g., the use of energy in \cite{MonroeRead2008} or the use of attractors in \cite{orr2014complex}. Prior work has considered attitudinal stability as a property of the network topology, expressed in the form of the magnitude of the edges on a graph \citep{MonroeRead2008} or the network structure itself \citep{OrrThrushPlaut2013,ShultzLepper1996}.  In principle, then, the CANAE approach is very closely aligned with prior constraint satisfaction computational models of attitudes networks. 

The deeper novelty of the CANAE framework lies in its use of constructs from network science and statistical physics to develop new kinds of theoretical predictions, e.g., in respect to the relation between the structure of the attitude network to its dynamics.  The natural path in such theoretical enterprises, to begin empirical testing of said predictions, has commenced \citep{DalegeMaas2019, DalegeMaas2017, DalegeVanderdoes2022, DalegeBorsboomvan2017,ZwickerDalege2020, ChambonDalege2022}.

Such advances in attitude theory come with subtle and nuanced theoretical conjectures.  For example, CANAE postulates that understanding the local topological characteristics of individual components/nodes in an attitude network should be sufficient to predict how perturbation of said components (e.g., via persuasion) will affect the global dynamics of attitude formation, what we dub as the \textit{extent of effect} of perturbing a component/node \citep{dalege2016,ChambonDalege2022,ZwickerDalege2020,DalegeBorsboomvan2017,DalegeMaas2017}.\footnote{This issue is of obvious practical import for clinical psychology \citep[see][]{Bringmann2018, Bringmann2021, bringmann2019centrality, WichersWigman2015, cramer2016major, burger2020bridging, haslbeck2021modeling} and other applied domains in which persuasion or behavior change are paramount for prevention, intervention and mitigation (e.g., public health \citep{OrrMortveitLebierePirolli2023,orr2014complex,orr2017galeabook,orr2017readbook}, climate change \citep{ThompsonClimate2023}, disaster preparedness \citep{SchlegelmilchCarlin2023}).}  Our theoretical apparatus should come to match in kind, a point we high-light next.
\subsection{Mathematical Analysis and Simulation are Complements}
In this section, we illustrate the potential for a complementary relationship between mathematical analysis and simulation in developing theory in attitude research.  In essence we ask: What is the nature of their respective contributions for providing quantitative, theoretical predictions that are amenable to empirical validation?  Using CANAE for illustration, we show that simulation and mathematical analysis, when considered together, can enrich our theoretical understanding of attitudinal phenomena.  But first, we must define the CANAE system and make clear some of the details of the CANAE theoretical claim that we have employed for our illustrative example.   
\subsubsection{The CANEA System}
The technical details of CANAE implementation are as follows:  

\begin{itemize}
\item There is a graph $G = G(V,E)$ consisting of a collection of beliefs (vertices from a set~$V$) and relations between them (weighted edges from a set~$E$).
\item The state of vertex $i\in V$ is $x_i \in K_i$ where $K_i$ is the state set for that vertex. 
\item For all $i$ we have $K_i \in \{-1,1\}$.
\item The system state is $x = (x_1, x_2, \ldots, x_n)$.
\item The system global energy $H$ is defined using all $i \in V$ by $H(x) = -[\sum_{i \in G} \tau_i x_i + \sum_{j\in N_G(i)} w_{ij} x_i x_j]$ where $N_G(i) \subset V$ is the set of neighbors of $i$ in $G$, \emph{not} including $i$, $w_{ij}$ is the weight of the edge $\{j,i\}$ and $\tau_i$ is the baseline parameter for vertex $i$ (we use $\theta_i$ interchangeably with $\tau_i$ throughout the text).  Assume that $w_{ij} = w_{ji}$.
\item For $i\in V$ let $\sigma_i \colon \prod_{i=1}^n K_i \longrightarrow \mathbb{R}$ be the function defined by $\sigma_i(x) = H(x) - H(\bar{x})$ where $x$ and $\bar{x}$ are configurations given the current and opposite state of vertex $i$, respectively.
\item For each vertex $i$ we define its vertex function as $\phi_i(x) = 1/(1+e^{-\sigma_i(x)/t})$ where $t$ is the temperature, a parameter
of the system; this co-determines the probability that at any point in time a vertex $i$ will flip to its opposite state: $P(c \longrightarrow o) = \phi_i(x)$. 
\end{itemize}

A typical instance of CANAE is a discrete-time, asynchronous simulation.  For each time step: (i) select a vertex $i$, (ii) compute $P(c \longrightarrow o) = \phi_i(x)$ and (iii) use $P(c \longrightarrow o)$ directly to decide if vertex $i$ will change its state.  Another common implementation is to draw $n$ samples of the system state $x$ from the Gibbs probability distribution (the Gibbs distribution is defined by the energe $H$ over the configuration space (all possible states)).  
\subsubsection{An Illustrative Theoretical Claim}
The debut CANAE article \citep{dalege2016} provided an estimate of an attitude network in reference to the U.S. 1984 presidential candidate Ronald Reagan using the American National Election Study (ANES) of 1984 \citep[see Figure 2, right-panel in][]{dalege2016}\footnote{The central aim of the debut CANAE article \citep{dalege2016} was to provide a measurement model of attitudes based on network principles, a point that we do not explore here. Our interest and purpose in using CANAE was in its interesting and subtle theoretical predictions.}. Their initial theoretical assessment of the CANAE system dynamics (an intuitive prediction, if you will) emphasized node attributes:  
\begin{displayquote}
How a given node is connected in the network will influence whether and how change in this node will spread to other nodes. \citep[][p. 10, paragraph 7]{dalege2016}
\end{displayquote}
\noindent
Two network attributes of a node were singled out as important: membership in a cluster and node centrality.  Two sets of nodes in the ANES-Reagan data were accompanied by exemplary node-level assessments:  the nodes representing Reagan as setting a good example and whether he cares about his constituents, each of which is shown in respective order in the following two quotes:   
\begin{displayquote}
Thus, whether change in the negative affect cluster would spread through the network would depend on whether you change your mind that Ronald Reagan sets a good example. \citep[][p. 11, paragraph 3]{dalege2016}
\end{displayquote}
\noindent
and,
\begin{displayquote}
For example, the evaluative reaction with both the highest degree and highest closeness in the network of the attitude toward Ronald Reagan is the judgment of whether he cares about people like oneself. It is thus likely that ... change in this judgement would affect the attitude network to a large extent. \citep[][p. 11, paragraph 5]{dalege2016}
\end{displayquote}

These theoretical assessments, or intuitive predictions, refer to the consequence or extent of effect of the perturbation of single, individual nodes:  a node's centrality and its cluster membership will relate to its extent of effect\footnote{A single node is the limiting, simplest case; the CANAE theory can accommodate multiple node perturbation.}.  Subsequently, this original, theoretical but informal prediction was put to various kinds of tests \citep{ZwickerDalege2020,ChambonDalege2022,DalegeBorsboomvan2017,DalegeMaas2017}, something we focus on next.   
\subsubsection{Complementarity In Action}
The simulation work on CANAE that serves best for our illustration--using similar data as in \citep{dalege2016}--comes from \cite{DalegeMaas2017}.  Other works in the CANAE canon, both empirical and simulation based, were considered for our example but lacked some features we desired for our illustrative case.  Of the studies that directly addressed the extent of effect of perturbation, some were correlational and thus did not offer the control required for a perturbation study \citep{ChambonDalege2022,ZwickerDalege2020,DalegeBorsboomvan2017}. Of these, one included simulation but did not sufficiently suit our needs because it lacked perturbation methods \citep{DalegeBorsboomvan2017}. Some work in the CANAE canon addressed other important issues, but did not broach the topic of vertex perturbation and extent of effect \citep[e.g., ][]{DalegeMaas2019,DalegeBorsboom_2018,DalegevanderMaas2020}.  

The principal result for our illustrative example in \cite{DalegeMaas2017} was a difference in the sum score, defined as $\Sigma_i x_i$ (see above for the definition of $x_i$; this is a system-level state of all vertices), between two conditions--one condition perturbed the most central vertex by forcing its $\tau_i$ to a value of 1; the other condition did the same for the least central vertex.  The difference in sum score was approximately 2 points; the mean of 1.18 (SD= 7.99) for most central vertex and a mean value of -1.04 (SD = 7.84) for the least central vertex.

These simulation results provided important initial insights into the extent of effect phenomenon (described above):  (i) plausibility--vertex centrality may in fact play a role in the extent to which perturbation will have an effect, (ii) theoretical predictions in respect to the direction of the effect (as was explored in \cite{ChambonDalege2022,ZwickerDalege2020,DalegeBorsboomvan2017}), (iii) a useful, clear methodological approach for further testing; the platform and data they used are open-source.  We might imagine that next steps--continuing with the simple, limiting case of perturbing only one node--would be towards more detailed predictions which (i) vary the parameters of the system extensively in conjunction with (ii) estimating a function-like relation between vertex centrality and extent of effect.  Such efforts could serve as more detailed theoretical predictions for future empirical validation.  

We now move on to the mathematical analysis of a idealized simple case, one that is directly relevant for the extent of effect phenomenon (described above):  the  stability analysis of an asynchronous Hopfield model with only one attractor state $\xi_i$ (one stored pattern or, if you will, one stored, stable attitude) in which the thresholds $\theta_i = 0$, the states $\in \{-1,1\}$,  $w_{i,j}$ restricted to $\in \{-1,1\}$ and the sign function is used for the node state dynamics\footnote{For this simple case we borrow notation from \citep{HertzKroghPalmer1991}}.  Borrowing from the mathematical results in \cite{HertzKroghPalmer1991},  we know that if the initial state $S_i$ is equivalent to the attractor state, $S_i = \xi_i$, then the system will likely not change.  Further, if more than 1/2 of the nodes are correct, meaning more than 1/2 of $S_i = \xi_i$, then the system will likely settle to $\xi_i$ as the system will correct the incorrect bits in $S_i$. In reference to the \citeauthor{dalege2016} case, flipping one bit from $S_i$ would have near zero effect on the other nodes if the system was in or near its attractor state $\xi_i$.  Notice that flipping only one bit (from state $\xi_i$) would nearly guarantee, if you let the system evolve, that the bit would flip back to its original state the next time it was updated with probability 1 and, since no other bits would change, the system would settle to $\xi_i$.   Finally, it is also true that if nearly 1/2 (or greater) of the bits were different from $\xi_i$, then the flipping of one bit $k$ would likely have an effect.  Notice that the latter case is the only case that predicts perturbing one node would make a difference in the system state.  

What can we learn from this idealized analytic case?  The effect of flipping a bit can depend on the state of the system:  When the system is in or near its attractor, flipping a bit will have a near zero extent of effect; and, the system must be far from its attractor for the flip of one bit to have an effect.

%
To recapitulate, our illustrative comparison addressed, from two angles, the problem of whether perturbing a node in the system will likely have an extensive effect on the system.  We were not out to resolve this research question, but to show the complimentary nature of simulation and mathematical analysis for such questions.  The simulation analysis justified the theoretical claim, at a basic level, that node centrality moderates the effects of perturbation of a node in the system.  The mathematical analysis, at an equally basic level, makes plain that attractors can be very useful in defining the general conditions under which perturbation of a node will have an effect, ceteris paribus.  Notice that the constraints of both approaches are also complementary:  Stability analysis, in its analytic, mathematical form, doesn't account for network structure and is thus mute on the node centrality issue.  In contrast, the simulation approach offered no guarantee on performance outside the parameters of simulation some of which, in this case, were derived from one empirical survey of one specific population.\footnote{The network structure is a parameterization of the simulation approach.}  Taken together, caveats aside, we have convergent, high-level insights from two formal techniques, each of which offers a different view of the problem.  More importantly, their joint consideration can afford dialog, if you will, between them.   

We offer a quote from a classic work in Ising-like systems \citep{HertzKroghPalmer1991} that captures the general notion of the complementarity we have attempted to illustrate:
\begin{displayquote}
... we are usually not satisfied with simply stating or deducing a given result, but instead try to show the reader how to think about it, how to handle and hold it. \citep[][p. XX, paragraph 2]{HertzKroghPalmer1991}
\end{displayquote}

We now move to a more general mathematical theory, Graph Dynamical Systems, one that has the flexibility to provide results for a wide range of complex attitudinal phenomena.
%
%
%
%
\section{A Mathematical Attitude Theory}
We now introduce our mathematical theory, Graph Dynamical Systems, to explore a prominent computational model of attitudes, the Attitudes as Constraint Satisfaction (ACS) model \citep{MonroeRead2008}. In outline form, we will provide: (i) the requisite technical details of Graph Dynamical Systems (the mathematical theory we use), (ii) a high-level summary review of the ACS model and its gaps in terms of making precise testable predictions, (iii) a sketch of how one might move from mathematical analysis to useful experimental predictions using the ACS model \citep{MonroeRead2008}.  
\subsubsection{Graph Dynamical Systems}\label{gds}
Although Graph Dynamical Systems (GDS) appears outside of the psychological context, it is well suited for understanding attitude dynamics; GDS was developed in the context of socio-technical systems writ large, but was meant as a general abstraction for modeling and analyzing the discrete dynamics of networked systems.

The mathematical and computational theory of GDS (see, e.g., \cite{Mortveit:01a,Mortveit:07,Goles:90,Rosenkrantz:15,Mortveit:23}) is largely concerned with finite state sets such as~$\{0,1\}$ and specific update mechanisms used to assemble local dynamics on agents\footnote{Agents map onto vertices or nodes of a graph.}
into global dynamics of the complete system. Formally, a sequence of vertex functions $(f_i)_i$ indexed by the agents will, by applying an update scheme $U$, assemble to a map~$F_U \colon K^n \longrightarrow K^n$ where $K$ is the state set of each agent. For example, for a parallel update scheme with $n$ agents/vertices, we have
\begin{equation}
\label{eq:gds1}
    F_U\bigl(x=(x_1, \ldots, x_n )\bigr) = \bigl(f_1(x), \ldots, f_n(x)\bigr) \;,
\end{equation}
where the function $f_i$ captures the behavior of vertex~$i$.
The variables which the functions $f_i$ consumes capture the dependencies among the corresponding agents; we encode these through the dependency graph~$G$. 
\textit{In terms of the present article, contemporary computational models of attitudes--e.g., Hopfield models, Ising-like models, fully recurrent neural networks--are special cases of GDS.}

Existing mathematical and computational theory of GDS deals with how structural properties of the functions $f_v$, properties of the network~$G$, and the choice of update mechanism translate into properties of the system, captured through the state space dynamics. All standard questions and topics from dynamical system theory such as stability and control are studied. 
For example, it is well known that binary threshold GDS under sequential update mechanisms (see, e.g.,~\cite{Mortveit:07,Goles:80})  have only fixed points as attractors and these are invariant with respect to the choice of update sequence~\citep{Mortveit:07}, while the parallel update method, it turns out that periodic orbits of length~2 can also manifest~\citep{Goles:90}. 

The examples we provide below mark a way of using GDS to build a rigorous foundation of attitudinal networks.  Our focus will address the Attitudes as Constraint Satisfaction \cite{MonroeRead2008} model.  The general form of this models is captured completely in the GDS formalism:
\begin{itemize}
\item There is a graph $G = G(V,E)$ consisting of a collection of beliefs (vertices from a set~$V$) and relations between them (weighted edges from a set~$E$).  (Social psychologists will be familiar with nodes and weights (vertices and relations) in an artificial neural network.)
\item Each vertex $i\in V$ is assigned a dynamic state $x_i \in K_i$ where $K_i$ is the state set for that vertex. Generally, we have $K_i = [a_i,b_i]\subset\mathbb{R}$ where $a_i$ and $b_i$ are bounds on the vertex state values.
\item The system state is $x = (x_1, x_2, \ldots, x_n)$.
\item The edges, which are directed, are defined by a real-valued matrix $W = [w_{ij}]$. An edge $e$ from vertex $i$ to vertex $j$ is written $e = (i,j)$ and has associated edge weight $w_{ij}$.
\item For each vertex $i\in V$ there is a function $\sigma_i \colon \prod_{i=1}^n K_i \longrightarrow \mathbb{R}$ performing a local computation for vertex $i$ that captures both vertex biases and some form of coupling with other vertices through the the matrix $W$, e.g., $\sigma_i(x) = \sum_{i\ne j} w_{ij} x_j$.
\item Finally, for each vertex $i$ there is a vertex function of the form $f_i = \phi_i \circ \sigma_i$ where $\phi_i \colon \mathbb{R} \longrightarrow \mathbb{R}$ is for instance a threshold function, like Heaviside.
\item These kinds of models are typically explored through discrete-time, asynchronous simulations where, for each time step, one selects a vertex $i$ and evaluates $f_i$, and  instantiates a state change only for vertex~$i$. 
\end{itemize}
\subsubsection{Attitudes as Constraint Satisfaction}\label{contemporarymodels}
The ACS model was developed to demonstrate, as proof-of-concept, that dynamic network models could capture some of the key empirical patterns in attitude research via simulation; no mathematical analysis was provided.  This model was not designed to capture real human attitudinal contexts or to capture a specific set of experimental data.  Instead, the ACS leveraged a high degree of abstraction and a low degree of specificity to offer a proof-of-concept model, one that would spur future development.  

The primary measure of the ACS model \citep{MonroeRead2008} was the dynamics of a single vertex, called the evaluative vertex ($x_e$) that served as the evaluation of the attitude object.  The graph was partitioned into two competitive substructures--one to represent knowledge related to the attitude object (e.g., the presidential candidate is kind and intelligent) and the other to capture persuasive attempts against the existing knowledge.  A typical simulation trained the model to gravitate toward a positive evaluation of the attitude object, i.e., a positive value of $x_e$.  After training, the model was probed and perturbed to test the effects, theoretically, of specific kinds of persuasion, reasoning (e.g., motivated reasoning), mere thought on polarization, and elaboration likelihood. Other measures were used to provide some rudimentary understanding of the operation of the system (e.g., the energy of the system as coherence; the average states of the vertices in different partitions). 

Across a series of simulation experiments, a set of experimental factors captured aspects of the system that mapped onto real-world conditions of interest and features of variability in the structure of the system, e.g., size of knowledge structure (number of beliefs associated with the attitude object), relations of partitions (degree of competition between persuasion and knowledge), strength of the persuasion attempt (number of persuasion vertices), and finally, a form of processing capacity limitation.  

The objective of the ACS model was to demonstrate that certain configurations of initial conditions (e.g., the distribution of weights in $W$), learning (which typically fortified the initial conditions), and parametric configurations of the factors (e.g., size of the network, structural changes, capacity) could, in principle, mimic the coarse-grained features of key experimental phenomena. 

To summarize, the set of ACS simulations were, by design, highly-idiosyncratic, post-hoc instantiations of highly-stylized constructions.  Rigorous methods were not employed, nor have they been since, that would characterize this model in a systematic way. Also by design, the proof-of-concept simulations did not yield quantitative predictions that were amenable to experimental test. 

\subsubsection{From Mathematical Analysis to Experimental Predictions}
%
In this section we will (i) provide the formal description of the specific GDS form we aim to use for the proposed demonstration, (ii) define precisely the ACS model as a GDS, (iii) provide a stylized, textbook-like example of an experimental prediction from a simple formulation of the ACS, (iv) demonstrate, by example, how we envision the formulation of theoretically important experimental predictions in the future.

A GDS we would consider for the ACS would be a \emph{weighted, block sequential graph dynamical system} over a set $V = \{1,2,\ldots,n\}$ that is constructed from a sequence of vertex functions $F = (f_i)_{i=1}^n$ and a map $U$ that for each time step $t\ge 0$
assigns a subset $U(t) \subset V$ whose states are to be updated at that time.
Given some initial system state $x(0)$, the dynamics of the system state $x(t) = (x_1(t), x_2(t), \ldots, x_n(t))$ is given by:
\begin{equation}
x_i(t+1) :=
\begin{cases}
    x_i(t)\;, & \text{if $i \not\in U(t)$} \\
    f_i\bigl(x(t)\bigr)\;, & \text{if $i \in U(t)$} 
\end{cases}
\end{equation}
The dependency graph $G$ associated to $F$ has vertex set $V$ and edges all $(i,j)$ for which $f_j$ depends non-trivially on $x_i$. The graph $G$ captures the possible interactions among vertices. We associate to $F$ the matrix $W = [w_{ij}]\in M_n(\mathbb{R})$ of edge weights; here it is assumed that $w_{ij} \ne 0$ if and and only if $(i,j)$ is an edge in $G$.

\begin{definition}
~\label{D:att_mem_sys}
We set $V=\{1,2,\ldots,n\}$ and specify the following:
\begin{itemize}
\item An \emph{object} vertex $v = 1$ with state $x_1\in \{0,1\}$,   capturing the absence/presence of an \emph{object} to be evaluated, 
and an \emph{evaluation} vertex $v = n$ with $x_n \in [-1, 1] \subset\mathbb{R}$. Here $x_n<0$ (resp. $x_n>0$) models a negative (resp. positive) attitude toward the object, with $|x_n|$ representing the \emph{strength} or \emph{degree of polarization} towards the object.
\item A partition of the remaining vertices $\{2,3,\ldots,n-1\}$ into non-empty subsets $C$ and $P$. The set $C$ is called the \emph{cognitive} partition and represents \emph{features, concepts or interior beliefs} held about the object, while $P$, the \emph{persuasion} partition, represents \emph{exterior persuasive influences} regarding the object.
\item Parallel update: for all time steps $t\ge 0$ we have $U(t) = V$. 
\item Vertex functions: let $e_1$ denote the unit vector $(1,0,\cdots,0)$, $W_i$ the $i^{\text{th}}$ row of $W$, and $\langle x, x' \rangle$ the inner product of vectors $x$ and $x'$. The vertex functions are defined by
\begin{equation*}
f_1(x) = \langle e_1,x\rangle \;,
\quad 
f_{1<i<n}(x) = 1 \bigr/({e^{-\langle W_i,x\rangle}+1}) \; ,
\text{\  and, \ }
f_n(x) = ({e^{\langle W_n,x\rangle}-1})\Bigr/({e^{\langle W_n,x\rangle}+1}) \;.
\end{equation*}
\item Stopping criterion: for $\theta\in\mathbb{R}$ the $\theta$-\emph{stopping time} $t^*$ is the smallest time step such that the norm 
$||x(t^*+1)-x(t*)||\le \theta$.
\end{itemize}
\end{definition}
%
%

Next we provide an example of predictions for a hypothetical experiment. We focus on what parameter regimes and weight ranges may give rise to successful or unsuccessful persuasion attempts. Although simple, this provides key elements of what we envision would apply to more advanced models addressed in the psychological literature.  

\begin{figure}[ht]
\centering{
\begin{tabular}{ |c|c| } 
\hline
$$
\begin{tikzcd}
 & p\arrow[d,"\delta",dashed] &\\
 & c\arrow[dr,"\alpha'"] &  \\
o\arrow[ur,"\alpha"] \arrow[rr,"\gamma"]& & e
\end{tikzcd}
$$
&
$$
\begin{tikzcd}
 & x_p=1\arrow[d,"\delta",dashed] &\\
 & x_c\arrow[dr,"\alpha'"] &  \\
x_o=1\arrow[ur,"\alpha"] \arrow[rr,"\gamma"]& & x_e \stackrel{?}{\lessgtr} 0
\end{tikzcd}
$$
\\
\hline
\end{tabular}
}
\caption{Left: the network of the basic example. Right: dynamics evolving over the network for state $(x_o = 1, x_p = 1, x_c, x_e)$ and assessing whether parameter choices cause compliant or non-compliant behaviors with the persuasion attempt ($x_e < 0$ or $x_e > 0$).}
\label{fig:ex1}
\end{figure}
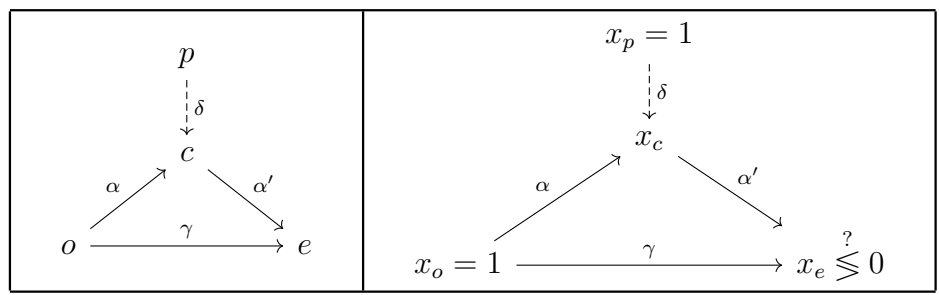

We use a system with four vertices $V = \{o, p, c, e\}$, where $o$ is the object, $e$ the evaluation, $c$ is for cognition, and $p$ is for persuasion. 
The weight matrix $W$ has four non-zero entries given by~$w_{oc}=\alpha$, $w_{ce}=\alpha'$, $w_{oe}=\gamma$, and $w_{pc}=\delta$ as shown on the left in Figure~\ref{fig:ex1}.
The states $x_o=1$ and $x_c=1$ represent the presence of the object and \emph{cognitive engagement} with the object, while $x_e>0$ represents a positive attitude towards the object. The $\alpha$ parameter represents \emph{active attention} towards the object while $\delta$ represents a competitive coupling to an exterior \emph{persuasive influence} represented by the state $x_p=1$. This attention competition is modeled by tuning the value of the parameter $\delta$ from~$0$ to~$1$. Finally, $\alpha'$ represents the \emph{cognitive contribution} to the attitude value of $e$ while $\gamma>0$ represents a \emph{automatic associative bias} towards the object. As the $\delta$-tuning takes place, we wish to study which scenarios (parameter settings) lead to \emph{compliance with persuasion} with the persuasive influence which seeks to change the $x_e>0$ (positive attitude) to $x_e<0$ (negative attitude).

\begin{figure}[ht]
\centerline{
\includegraphics[width=0.5\textwidth]{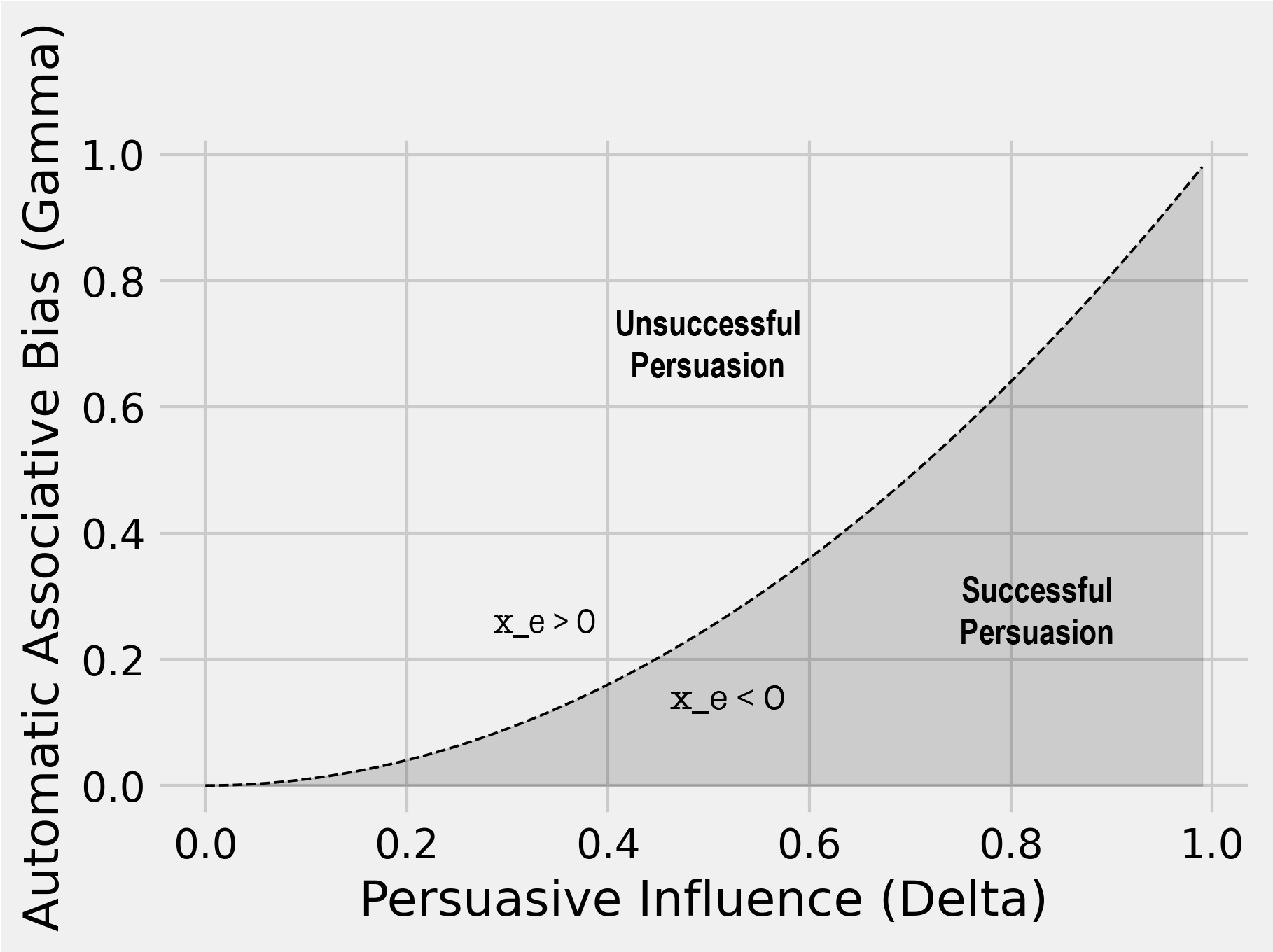}
}
\caption{A simple, textbook-like example of experimental predictions via mathematical analysis for the Attitudes as Constraint Satisfaction (ACS) model \citep[]{MonroeRead2008}. We have introduce the two parameter relation $\alpha'=f-\delta^2$ and assert that $\alpha + \delta = 1$.  Under this particular choice, we obtain the boundary curve $\gamma = \delta^2$ separating the non persuadable (light; $x_e < 0$)) and the persuadable regions (dark gray; $x_e > 0$) as a function of manipulations of persuasive influence $delta$ and automatic associative bias $gamma$. See text for details.}
\label{fig:c_domain}
\end{figure}

In our case, the vertex functions of the ACS model for vertices $c$ and $e$ are given by $f_c(x) = \alpha x_o+\delta x_p$, and $f_e(x) = \alpha' x_c + \gamma x_o$ with the remaining two being constant functions $f_o(x) = 1$ and $f_p(x) = 1$.
We note that any initial state~$x(0) = (x_o = 1, x_e=1, x_c=1, x_p = 1)$ is eventually mapped onto the fixed point 
\begin{equation}
\label{eq:acs_basic}
x_o = 1, \quad 
x_p = 1, \quad 
x_c = \alpha + \delta, \quad\text{and}\quad
x_e = \alpha'( \alpha + \delta) + \gamma\;.
\end{equation}
The boundary in parameter space separating successful and unsuccessful persuasion can be obtained as the manifold defined by equating $x_e$ in~Equation~\eqref{eq:acs_basic} to~0, that is, $\gamma = -\alpha' (\alpha + \delta)$. 
\emph{Here is the key insight: the expression for $x_e$ allows one to (a) identify which parameters to target in an experiment, and to (b) quantify the magnitude of adjustments to the chosen parameter(s) in order to obtain a specific outcome.}
With a model having many parameters, one may want to restrict this space by introducing relations among them. In this example, we relate $\alpha'$ and $\delta$ through the function $f$ as  $\alpha'=f(-\delta^2)$.  If we control $x_c$ to be 1 (via $\alpha + \gamma = 1$) , we can derive $\gamma = \delta^2$ to understand the relation between the degree of persuasion and the degree of automatic associative bias.  Figure \ref{fig:c_domain} illustrates an experimental prediction in terms of when a persuasion attempt would be successful or not.
In practice, one would relate parameters and possible constraints to experimental mechanisms and controls.

It is important to revisit the purpose of this exercise, a simple, textbook-like example of the process from mathematical formulation to experimental prediction in attitudinal networks. It was not to show the nature of the kinds of predictions we envision for future work (we discuss this below) but to show what we mean by making precise empirical predictions on dynamic networks.  

Next we provide a sketch of what formal methods might look like in a more realistic setting using the ACS model.  We focus on a question of direct interest to attitudes.
This question has two parts and can be broadly categorized as having directly to do with the act of persuasion:
\emph{What is the ratio of persuasion elements to cognitive elements for a successful persuasion, and is this ratio an invariant with respect to any characteristics of the act of persuasion?}

To formalize this (using some reassignment of parameters in comparison to the above example), we let $M=(W,F,U)$ be the GDS formulation of the ACS model previously described. The ratio of interest, call it $Q$, then naturally maps in our formulation to 
$$
Q(M):=\frac{|C|}{|P|}.
$$
We can control for the size of the total system, which means fixing $n=\dim(W)$. We let $|P|:=\alpha (n-2)$ where $\alpha\in [0,1]$ represents the fraction of nodes in the system that are persuasive. We label $M_\alpha:=(W(\alpha),F(\alpha),U(\alpha))$ to denote this interpretation. Since $|C|+|P|=n-2$ we have in terms of $\alpha$
$$
Q(M_\alpha):=\frac{n-2}{|P|}-1= \frac{n-2}{\alpha(n-2)}-1=\frac{1-\alpha}{\alpha}.
$$
Thus, for a system of fixed size $n$, any $x(0)\in\mathbb{R}^n$ initial state now has an interpretation of an  $\alpha$-fraction of its coordinates belonging to persuasive elements. We denote $x(t^*_\alpha):=M_\alpha[x(0)]$ the state of the system $M_\alpha$ at its $\theta$-stopping time $t^*_\alpha$, for the fixed initial condition $x(0)$. We can then propose to study variational statements of the form
$$
|\alpha-\alpha'|\le \epsilon,\quad \implies |x_n(t^*_\alpha)-x_n(t^*_{\alpha'})|\le \delta,
$$
or more generally ,
$$
|\alpha-\alpha'|\le \epsilon,\quad \implies ||x(t^*_\alpha)-x(t^*_{\alpha'})||\le\delta,
$$
where $||\cdot||$ denotes suitably chosen matrix/vector norms, and $\delta\in\mathbb{R}$ represent the errors in evaluation and in terminal system state respectively, both taken under $\alpha$-perturbations controlled by $\epsilon\in\mathbb{R}$.

The implications formulated above have natural interpretations which will be of interest. For instance, a parameter pair $(\epsilon,\delta)$ satisfying the first implication says that a variation of at most $\epsilon$ from the current persuasive strength $\alpha$, would guarantee a change of no more than $\delta$ in the attitude the system exhibits under the constant initial circumstances of $x(0)$. Namely, all things being equal, we can relax/increase our persuasion by no more than $\epsilon$ and expect a drift of no more than $\delta$ in attitude. 

We formulate the question in this manner as it conveniently allows us to see the problem in terms of a continuous parameter $\alpha$ (a proxy for the proportion of persuasive nodes). Using this formulation this naturally becomes a question about the continuity of the dynamics of the system when $\alpha$ varies. For instance for a fixed $\alpha$, if the collection of vertex local functions is interpreted as a phase space function that is iterated under block sequential update, and the local vertex functions happen to be such that $F$ is contractive, then this becomes an iterated function system. Then, by~\cite{Hut:81} we know a unique fixed set exists for such a system which is now $\alpha$-parametrized. The "volumes" of these sets then bound the attitude changes possible for any initial condition state chosen from such a set, and this bound can change as $\alpha$ is varied. 

This more realistic example provides a sketch of one way to expand formal methods to more realistic psychological networks.  The key takeaway point is that, by example, persuasion attempts may be modeled by the ratio of cognitive vertices to persuasion vertices. Using this formulation, we could consider variational methods to make experimental predictions about the degree of change in the attitude network given a specific degree of persuasion.  

Using variational and other formal methods, it may be feasible to develop very nuanced hypotheses related to a broad set of research questions per the ACS:\footnote{Keep in mind that vertex and vertices correspond to specific beliefs or facts in the cognitive partition of the ACS and to persuasive statements in persuasion partition of the ACS.}  
\begin{itemize}
\item Is it possible to identify vertices that would be more effective for persuasion attempts.  For example, assume we are limited to only perturbing 10\% of the persuasion vertices, under what conditions should we (should we not) target specific vertices. Assuming such conditions do exist, what defines the heterogeneity in effectiveness across vertices (e.g., centrality of vertices)?
\item Assume some vertices of the cognition partition are not measured.  How does this affect our understanding of persuasion and learning?
\item Assume that the set of vertices that are measured includes both relevant and irrelevant vertices in the cognition partition. How does this affect our understanding of persuasion and learning?  Can we model the irrelevant vertices as noise?
\item What is the effect of precision in estimation of edge values on the dynamics of the system? Are binary values sufficient, especially as networks get bigger? Or do we need more precise/granular measurement? What level of precision do we need?
\item How do we represent/capture learning in the model? Change in vertex strength? Change in edge strength? 
\item Can we identify points of potential maximal change or vulnerability using the wide range of network characteristics that have been developed.  How would changes in those characteristics influence dynamics?
\end{itemize}


    
\section{Conclusions}
    At its core, this article proposes a mathematical theory for attitudinal theory.  Whatever does this mean?  Much of the theoretical apparatus (theory) in social psychology, and to some non-negligible degree in psychology as a whole, is largely divorced from what qualifies as theory in other disciplines.  In the physical sciences and engineering, for example, theory is a reference to a circumscribed set of mathematical results (conjectures and proofs) stemming from a set of theory-specific axioms (e.g., graph theory). Physics is canonical in this respect; its largely math, so many physical theories are defined by the math.  In engineering and in computing, references to theory are nearly exclusively in reference to mathematical theory (e.g., theory of computation is a branch of discrete math).  So, the essence of our article is its attempt to provide a similar foundation for attitude research.  We hope, as attitude research matures, that any given reference to theory may at a future point in time refer to a mathematical theory, graph dynamical systems being just one possible example\footnote{We surmise that mathematical theory might, for example, help to address the dynamics issue outlined in recent work in clinical psychology \citep[see][]{Bringmann2018, Bringmann2021, bringmann2019centrality, WichersWigman2015, cramer2016major, burger2020bridging, haslbeck2021modeling}.}.

In closing, we frame our work in reference to one of the central scientific issues in contemporary scientific psychology:  the so-called replication crisis \citep[see][]{Nosek-2022}.  Our work exemplifies an unsung heroine of the replication crisis: formal computational and mathematical modeling.  In contrast to the oft-sung heroes of this crisis--better data, better statistical methods and deeper administrative controls (e.g., pre-registration)--computational and mathematical modeling are of limited repute, largely due to the degree to which they are misunderstood in terms of use and value. A small, nascent effort to mend these rifts has erected a motley set of arguments for the necessity of formal modeling efforts in scientific psychology \citep[e.g.,][]{Oberauer-Lewandowsky-2019,Smaldino-2020,Robinaugh-2019,Fried-2020}, something that is aimed, by necessity, more towards social and clinical psychology.  The cognitive sciences, neural sciences and perceptual sciences use and train with such models with more regularity. Note, however, that there is an argument for all of scientific psychology, including these latter sub-disciplines, to persist in driving towards the development of a mature science, one with a cumulative, systematic march to overarching theoretical clarity \citep[see][]{Muthukrishna-Henrich-2019}.  Our work presented here hopes to contribute towards this goal in social psychology.

    
\section{Author Note}
The work in this study was funded by NSF Grant Nos.: 1520359, 2002626 and 2200112. 

\bibliography{example.bib}

\end{document}